# Augmented reality in higher education: a case study in medical education

Danai Korre[1,2] and Andrew Sherlock[1,3]

[1] EdAR, Edinburgh, UK
[2] The University of Edinburgh, Edinburgh, UK
[3] Strathclyde University, Glasgow, UK
d.korre@ed.ac.uk

**Abstract.** During lockdown, we piloted a variety of augmented reality (AR) experiences in collaboration with subject matter experts from different fields aiming at creating remote teaching and training experiences. In this paper, we present a case study on how AR can be used as a teaching aid for medical education with pertinent focus on remote and social distanced learning. We describe the process of creating an AR experience that can enhance the knowledge and understanding of anatomy for medical students. The Anatomy Experience is an AR enhanced learning experience developed in collaboration with the Medical School of the University of Edinburgh aiming to assist medical students understand the complex geometry of different parts of the human body. After conducting a focus group study with medical students, trainees, and trainers, we received very positive feedback on the Anatomy Experience and its effects on understanding anatomy, enriching the learning process, and using it as a tool for anatomy teaching.

## 1 Introduction

The resurgence of extended reality applications in recent years created a renewed research interest in educational augmented and mixed reality (AR/MR). It has been shown that augmented reality (AR) can enhance teaching more than a textbook [1] due to its interactive nature, enhanced visualisation, and immersion. The advancing technology of AR allows for interactions between real objects and virtual objects thus allowing for a better understanding of complex geometries [2],[3],[4]. By using AR, students can experience and interact with objects or procedures in detail, by simply using a smartphone, tablet, or AR glasses. This makes AR highly effective when teaching geometric and spatial concepts, often found in science, technology, engineering, and math. An enhanced learning experience based on AR has been developed in collaboration with the Medical School of the University of Edinburgh (UoE) and is referred to as the Anatomy Experience. The aim of this experience is to help teach students the orientation and position of X-rays for complex parts of human anatomy such as the pelvis. The physical three-dimensional (3D) model of the pelvis can be examined and with the help of AR the user can control the axial, coronal and sagittal planes that are used to transect the body and see how the corresponding X-rays look like.

### 1.1 Requirements and scope of the Anatomy Experience

Several meetings have taken place between EdAR and the University of Edinburgh's Medical School to discuss the specifications and possible applications of AR in medical higher education. It has been considered that human anatomy would be a potential application area because of its complex geometry combined with the ability of AR-based learning to enhance the comprehension of complex information and simplify its delivery [5].

By following the EdAR process of developing experiences, we initiated a consultation with a subject matter expert (SME) to initiate collaboration, set a vision and build a time plan. We then collected the requirements, analysed them, and planned the next steps. By using a custom structured learning objectives questionnaire for collecting valuable data on subject matter expertise, we defined the areas that the experience will cover, important features, insights, requirements, learning objectives, aims, constraints and technical needs. By using these data in collaboration with the SME, the instructional designer generated an instructional needs analysis, and structured the content outline.

Table 1. Project overview summary

| | |
|---|---|
| **Objectives:** | Help teach students the orientation and position of X-rays for complex parts of human anatomy such as the pelvis |
| **Audience:** | Undergraduate medical students |
| **Constraints (time, place etc.):** | Online mostly, more likely for remote teaching |
| **What does the audience need?** | Access to a mobile device |
| **Synchronous or asynchronous?** | Can be used for both |

## 1.2 Training Objective

**Skills to be gained:**

- Understand the orientation and position of X-rays for complex parts of human anatomy such as the pelvis.
- Identify where the computed tomography (CT) image corresponds to the pelvic anatomy.

**Concepts to be covered:**

- Anatomy of human pelvis
- CT scan imaging

**Outcomes:**

By the end of this Experience students should be able to recognise the orientation and position of X-rays for complex parts of human anatomy and be able to locate where in the actual anatomy the imaging corresponds.

## 2 Design and development

After collecting the necessary data, the development and interaction design team generated the functional design of the experience. We also made user interface decisions and created a prototype. At the development stage, the SME feedback was integrated, and development commenced. The necessary content was identified and authored while the definition of the experience was developed in JavaScript Object Notation (JSON). The 3D printed parts and quick response (QR) cards were designed and generated at this point. A demo of the experience was presented to the SME for feedback. There are two sizes for the pelvis's physical models, a large and a mini which can be used by the student (Figure 1). The large model is intended for use in person in the classroom and the small model can be shipped to the student so they can practice on their own time and environment. There is also a version without the physical model where the pelvis is displayed digitally.

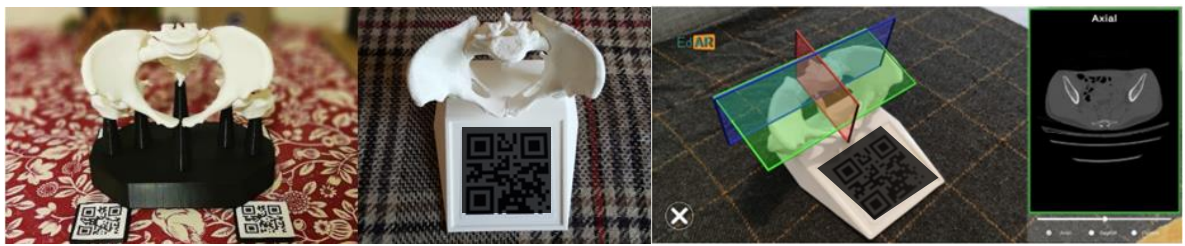

**Fig. 1.** On the left is the large 3D printed model of the pelvis. In the middle is the small 3D printed model of the pelvis. Image on the right is a screenshot of the AR-experience alpha prototype as seen on a smartphone screen.

## 3      Testing and integration

COVID 19 lockdown restrictions limited our access to sites and students which affected the development of the Experience, mainly the testing part of our development life cycle. We were however able to run a session with students during the UoE Clinical Skills Workshop where we piloted our Anatomy Experience (Figure 2). The cohort consisted of 19 undergraduate medical students, three postgraduate trainees and two junior trainers (non - consultant) who demoed the Anatomy Experience using our 3D printed models (Figure 1). The survey was conducted in English. The survey included questions about their impressions of the application for educational purposes and overall experience. The purpose of this survey was to gather some preliminary data and first impressions about the Anatomy Experience. The research results are intended to be used as a basis to update the existing application. Closed questions were used followed by a range of pre-coded responses apart from the comments section. Based on the responses, most participants strongly agreed that the application could enrich the learning process, improve their understanding of pelvic anatomy, and assist in the teaching of pelvic anatomy. Overall, the reception of the Anatomy Experience by the medical students, trainees and trainers has been very positive with a plethora of suggestions for expanding to other parts of the anatomy such as the knee, shoulder, and tissues.

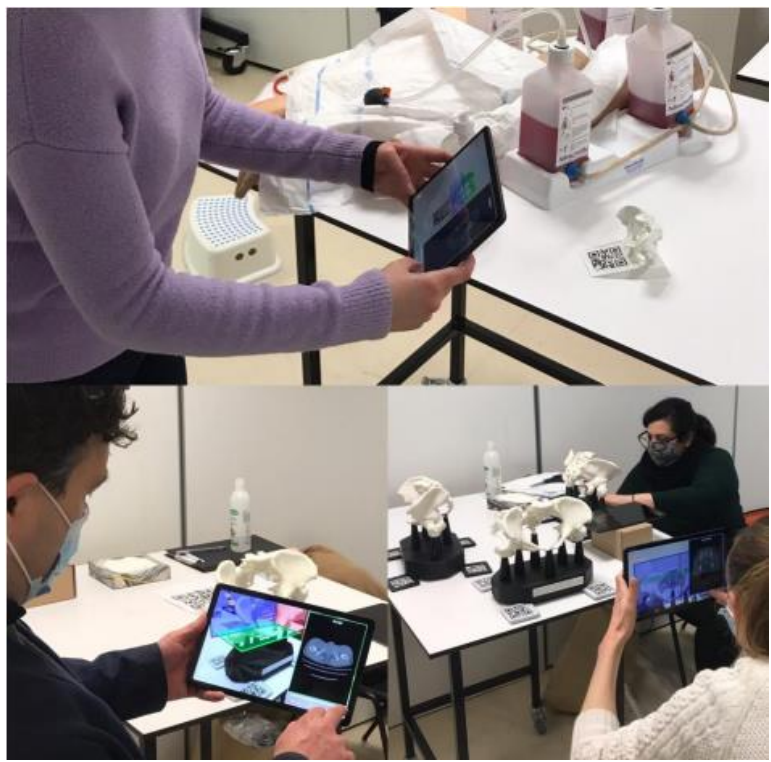

**Fig. 2.** Medical students and EdAR members trying out the Anatomy Experience

## 4      Challenges, summary, and conclusions

Screen estate is one of the challenges and, therefore, we launched a technographic survey collecting data from current medical students to design responsive user interfaces and exploit the screen size better. Another challenge with the proposed experience is that the 3D printed pelvises must be shipped to each student.
Educational AR has grown significantly in the last few years. The application we built combines the benefits of AR with the physical object of the pelvis, thus creating a mixed reality experience. Having a physical object was deemed to be particularly useful when teaching human anatomy both in the focus group and a workshop we run demonstrating the experience [6]. By creating small and cost-effective 3D prints of the pelvis that can be sent out to each student makes the Anatomy Experience ideal for remote teaching; this goes beyond the traditional media commonly used such as textbooks, video, and video calls.

## Acknowledgements

We would like to thank Andrew Hall, clinical teacher and PhD research fellow, for his assistance during testing. EdAR is a spin out company resulted from a European Institute of Innovation and Technology (EIT) digital funded project.